\documentstyle[11pt,newpasp,twoside,epsf]{article}

\def\edcomment#1{\iffalse\marginpar{\raggedright\sl#1\/}\else\relax\fi}
\reversemarginpar

\begin{document}
\title{Terrestrial Planet Finding with a Visible Light Coronagraph}
\author{Marc J. Kuchner\altaffilmark{1}}
\affil{Harvard-Smithsonian Center for Astrophysics, Mail Stop 20, 60 
Garden St., Cambridge, MA 02138}
\altaffiltext{1}{Michelson Postdoctoral Fellow}
\author{David N. Spergel}
\affil{Princeton University Observatory, Peyton Hall, Princeton, NJ 08544}

\begin{abstract}
Directly imaging extrasolar planets using a monolithic optical
telescope avoids many pitfalls of space interferometry and opens up
the prospect of visible light studies of extrasolar planetary systems.
Future astronomical missions may require interferometry for high
spatial resolution, but given that the first direct imaging missions
will probably fit into a single launch vehicle, the astrophysics of
planet finding calls for a visible light coronagraph as the first
space mission to search for extrasolar terrestrial planets.  New
coronagraphic techniques place the necessary dynamic range within
reach for detecting planets in reflected starlight.

\end{abstract}

\section{Introduction}

Bracewell's (1978) paper about using interferometry to separate
the light from an extrasolar planet from the light of its host star
spurred decades of research into interferometry as the canonical way
to detect extrasolar terrestrial planets.  Guided by NASA and ESA, a
cohort of research groups around the world has been attacking the
problems associated with interferometric planet finding: building
achromatic nulling beam combiners and cryogenic coolers, operating
formation flying spacecraft and automated fringe trackers, surveying
exozodiacal dust backgrounds, etc.  Laboratory tests suggest that a
mid-infrared (mid-IR) interferometer with nulling beam combiners can generate
the high dynamic range necessary for planet detection (Serabyn et
al. 1999), and surely one day a mid-IR interferometer will obtain the
accuracy required for extrasolar terrestrial planet finding.

But a monolithic optical telescope can also directly detect extrasolar planets.
New coronagraphic techniques using masks and stops and deformable mirrors
can potentially generate enough dynamic range to directly detect extrasolar
terrestrial planets with a single optical telescope.  We discuss
astrophysical reasons for preferring a visible-light
coronagraph for a first Terrestrial Planet Finder (TPF) and some
of the new optical techniques that can power it.

\section{Reasons for a Coronagraphic TPF}

\subsection{Inner Working Angle}

TPF must find an extrasolar terrestrial planet and take its spectrum.
No one knows how many stars must be searched to find such a planet,
but clearly, more is better.  A planet must have a minimum angular separation
from its host star to be detectable by a given TPF design.
This separation is the design's ``inner working angle''.
The smaller the inner working angle of a TPF design, the more distant
stars---and the more stars---it can search.

A coronagraph has an inner working angle between $2 \lambda/D$ and $4
\lambda/ D$ (pushing this limit calls for more research).
The primary mirror has longest dimension $D$, and visible
light TPF operates at wavelengths $\lambda=$0.4-1.1~$\mu$m.  For
example, a 4~m optical telescope would have an inner working angle of
$\sim 100$ milliarcseconds ($4 \lambda/D$ at $\lambda=0.5$~$\mu$m), or
1 AU at 10 pc.  Conventional optical telescopes typically have
circular primary mirrors, but since $D$ is the longest dimension of
the primary, elliptical mirrors are better for TPF. A coronagraph with
a 6~m $\times$ 2.7~m primary mirror, would be able to search roughly
three times as many stars as the 4~m example, given the same
collecting area.

A coronagraph offers a smaller inner working angle than an
interferometer given the same size optics.  A single-baseline nulling
interferometer has an inner working angle of $\lambda/(4 B)$, where
$B$ is the baseline and $\lambda$ is roughly 8.5--15~$\mu$m.  At first
glance, it might seem that at a given inner working angle, $B \approx
D$ within a factor of two.  However, the single-baseline nuller
generates a fringe pattern which varies as $\theta^2$ at the null,
where $\theta$ is the angle from the
optical axis, so $10^{-4}$ of the light from a solar type
star at 10 pc leaks through because of the finite size of the stellar disk.
This leak would overwhelm the signal from an Earthlike planet, which
would be $\sim 10^{7}$ times fainter than the star in the mid-IR.
A visible-light coronagraph produces the equivalent of
a $\theta^4$ null (at least) and further separates starlight from the
planet search area by imaging the star.  Nulling interferometers must
add additional long baselines to deepen the null, so that a 20~m
mid-IR interferometer TPF and a 4~m diameter optical coronagraph TPF
have comparable inner working angles.

\subsection{Planet Characterization}

After finding a planet, TPF must be able to characterize it---to
monitor it and measure its spectrum to determine whether or not it is
likely to be habitable.  We would like to eventually have both mid-IR
and visible wavelength ranges available to study extrasolar planet
atmospheres.  However, as a first measurement, a visible spectrum
would provide compelling science, easily explained to taxpayers.

Studies of how to characterize extrasolar planets generally use the
Earth as a reference point (Des Marais et al. 2001; Woolf et
al. 2002).  In the visible TPF band (0.4-1.1 $\mu$m), we can measure
the key biomarkers O${}_{3}$ and O${}_{2}$ on an Earth analog.  We can
also measure H${}_{2}$O on an Earth analog, and other biologically
important molecules CH${}_{4}$ and CO${}_{2}$ on an analog of the
early Earth.

Rayleigh scattering dominates the short-wavelength end of the Earth's
visible spectrum, providing the blueness of the sky.  Measuring this
phenomenon on an extrasolar planet would indicate the total column
depth of its atmosphere.
Visible light spectroscopy of the Earth also
detects a tantalizing signature from vegetation, slightly too
faint to detect on a true Earth twin, the ``red edge'' at 0.7~$\mu$m
(Seager \& Ford 2003).  TPF may not find little green men---but
visible light TPF could find big red plants!
Cloud patterns on the Earth have much higher contrast in the visible,
and they remain stable long enough that a time series of visible
photometry would indicate a planet's rotational period (Ford et al. 2001).

Mid-IR spectroscopy of the Earth can also measure O${}_{3}$,
H${}_{2}$O, CO${}_{2}$, and if the wavelength
range is extended shortward to 7~$\mu$m, possibly CH${}_{4}$.  The mid-IR
continuum also indicates a sort of mean temperature of the planet's
surface and cloud layers.  The mid-IR offers no direct O${}_{2}$
signature.  The main mid-IR biomarker, the O${}_{3}$ band
at 9.7~$\mu$m, is a particularly
strong feature and a sensitive indicator of small amounts of
O${}_{2}$.  But for the same reason, it is highly saturated in
the Earth's spectrum, making it a poor guide to the amount of
atmospheric oxygen on an analog of today's Earth.

\subsection{Exozodiacal Dust}

The brightest circumstellar source in the habitable zone of our
solar system is not a planet but a disk of zodiacal dust, amounting
to the equivalent of a single $\sim 50$~km asteroid crushed into
1--100~$\mu$m grains.
Other main sequence stars host similar
``exozodiacal'' clouds with up to $\sim 10,000$ times the surface brightness.
An IDL routine to compute the surface brightness
of exozodiacal clouds analogous to the solar zodiacal cloud
is publicly available at
http://cfa-www.harvard.edu/$\sim$mkuchner/.

An interferometric TPF acts like an antenna; it sums up the energy
collected over a broad swath of the sky, modulated by a fringe
pattern.  A mid-IR interferometer with a baseline of less than a
hundred meters can use a chopping scheme involving three or more
dishes to subtract the background emission from a smooth exozodiacal
cloud.  However such a scheme can not subtract the photon noise in the
solar zodiacal cloud or exozodiacal clouds.  This constraint drives
interferometric TPF designs toward using large 3 and 4~m diameter
mirrors.  An optical telescope, on the other hand, forms images,
directing most of the light from an exozodiacal cloud onto a different
detector pixel than the planet signal.

To complicate matters, known circumstellar clouds around other main
sequence stars are not smooth; they contain knots of emission, which
to a mid-IR TPF would resemble planets.  Taking a spectrum of a blob
of mid-IR emission---at a cost of perhaps weeks of integration
time---can distinguish dust clump false alarms from planets.  These
blobs may arise from the dynamical perturbations of planets embedded
in the dust.  To some degree, we can also decode images of dust
structures to infer the mass, orbital semi-major axis, and
eccentricity of a perturbing planet (Kuchner \& Holman 2003), and SIM
should eventually be able to warn us about perturbing planets, possibly
down to 3 Earth masses.  But exozodiacal dust will make interpreting
mid-IR interferograms of planetary systems more difficult and
ambiguous.

At a given inner working angle, the angular resolution of a visible
light coronagraph TPF (2--4 $\lambda/D$ for an image mask,
3--4 $\lambda/D$ for a pupil mask) is a few times finer than the angular
resolution of a mid-IR interferometer TPF ($\sim\lambda/L$ for a
typical 3 or 4 dish system, where $L$ is the total length of the structure).
This additional resolution allows a coronagraphic TPF to resolve structure
in an exozodiacal cloud that would confuse a similarly-priced interferometric TPF.
The linear resolution of a coronagraphic TPF is a factor of 2--4 higher
than the linear resolution of an interferometer TPF at a given inner
working angle; the improvement in exozodiacal background rejection
goes as roughly the square of this number.

\section{Very High Dynamic Range Optics}

The Earth is $\sim 2 \times 10^{-10}$ as bright as the sun at visible
wavelengths, where it shines in reflected sunlight.  Before we can
enjoy the advantages of working in the visible, we must show that
coronagraphs can suppress the light from a planet's host star to near
this contrast level---three orders of magnitude more contrast than
mid-IR TPF requires.  The sources of background light that limit the
contrast of a coronagraph divide into two categories: diffracted light
from pupil and mask edges, and scattered light from imperfections in
the optical surfaces.

\subsection{Diffracted Light}

In an ideal conventional optical telescope, such a faint source near a
bright star would be overcome first by the sidelobes of the point
spread function---Airy rings in the case of a circular aperture.  If
we ignore flaws in the optical surfaces, managing these sidelobes
amounts to controlling how the telescope diffracts light.

The ``diffracted light problem'' permits an infinite
family of solutions; this ``problem'' has been solved several times over.
These solutions generally resemble interferometers mathematically.
A coronagraph using a pupil-plane mask can be pictured as recreating
the $(u,v)$ coverage of a conventional interferometer with a carefully
tapered beam, with greatly reduced sidelobes (Spergel 2001, Kasdin et 
al. 2003, Guyon 2003).
A coronagraph using an image-plane mask recreates the fringe pattern 
of a nulling
interferometer (Kuchner \& Traub 2002), reducing both the sidelobes 
and the central peak
of the stellar image.  Both techniques offer a variety of options for
handling the necessary compromise between search area, inner working angle,
and throughput.  The two techniques can work together in concert, though
this possibility requires further exploration.

Coronagraphic masks draw on the class of ``band-limited'' functions
and their Fourier transforms.  The image mask transmissivity should be
a band-limited function; the pupil mask transmissivity must be the
Fourier transform of a band-limited function.  A band-limited function
has power only at low spatial frequencies. For example, a telescope
has a finite diameter, $D$, so a pupil mask can not use baselines
larger than $D$.  Likewise, a band-limited image mask cannot use
spatial frequencies higher than some fraction of $D$---the fraction of
the pupil diameter blocked by the Lyot stop.

The trick is to design an efficient system that is robust to errors in
manufacturing and control.  The image-plane and pupil-plane masks
described above can both be realized as binary masks, masks whose
transmissivities are either 0 or 1.  Binary masks can be created
simply by cutting holes in opaque material, to a tolerance of roughly
1/3000 of the diffraction scale.  Binary pupil masks can be
manufactured to this specification at a cost of $<$ \$10,000.  For
binary image masks (Kuchner \& Spergel 2003), the tolerance amounts to
roughly 20~nm.  This tolerance is routine for e-beam nano-lithography,
though such an image mask has not yet been constructed.

Controlling diffracted light using image masks rather than pupil masks
can provide somewhat better efficiency, particularly in terms of
search time, but an image-plane coronagraph is substantially less
tolerant than a pupil plane coronagraph to large-scale wavefront
errors, such as pointing.  The pointing requirements for an
image-plane coronagraph are identical to the pointing requirements for
a nulling interferometer with a quartic null given the same inner
working angle; the distribution of pointing errors can have standard
deviation up to $\sigma = 1.5$~mas for a 4~m coronagraph.  A
coronagraphic TPF could provide a choice of different kinds of both
pupil and image masks on a pair of filter wheels.

\subsection{Scattered Light}

Any telescope has figure and reflectivity errors in its optical
surfaces which scatter starlight into a background of speckles
throughout the image-plane, coronagraphic masks notwithstanding.  A
TPF coronagraph would manage these speckles using active optics.  The
TPF system must be stable enough that the corrections need only be
adjusted once every few hours, since it would take more than an hour
of integration time to acquire enough photons to measure the speckles
and decide how to update the correction.

Brown and Burrows (1990) referred to the ratio of the peak of the
expected planet image to the typical scattered light background by the
letter $Q$.  Some designs call for operating at $Q << 1$ under the
assumption that the speckle noise that will dominate their data can be
averaged out.  We believe that a TPF design should operate at $Q \ge
1$ to avoid the danger of systematic errors in the data and wavefront
control which cannot be averaged away.

Of course, the speckle need only be controlled over the planet search
area: angles $\sim 3 \lambda/D$ to $\sim 60 \lambda/D$ away from the
image of the star.  This search area corresponds to a range of spatial
frequencies on the primary mirror.  Achieving $Q \sim 1$ requires
wavefront errors over these mid-spatial frequencies of $< 0.5$ \AA\
rms in phase, and $< 10^{-3}$ rms in amplitude.

Figure~1 illustrates how a $Q \approx 1$ coronagraphic TPF image of a
planet might appear.  It shows the results of a broad band (0.66--1.0
$\mu$m) simulation of the performance of a 4~m coronagraph using a
linear binary image mask (Kuchner \& Spergel 2003) given 0.5 \AA\
rms errors in wavefront phase and $10^{-3}$ rms errors in
wavefront amplitude over the mid-spatial frequencies.  The pointing
error is assumed to be distributed in a Gaussian about zero with a
standard deviation of 1.5 mas, and the image mask has errors in its
shape at the level of 20~nm rms over spatial frequencies less than
the diffraction scale.  The grey curves show the images of a planet $2
\times 10^{-10}$ times as bright as its host star and an exozodiacal
cloud just like the solar zodiacal cloud at 5 pc assuming a 4m
diameter primary.  The top frame shows the attenuation of the planet
light by the mask, and the bottom frame shows the surface brightness
of the simulated images.  PSF subtraction techniques like spectral
deconvolution (Sparks \& Ford 2002) could efficiently extract the
planet signal from among the speckles.

\begin{figure}
\vspace{-2.0in}
\plotone{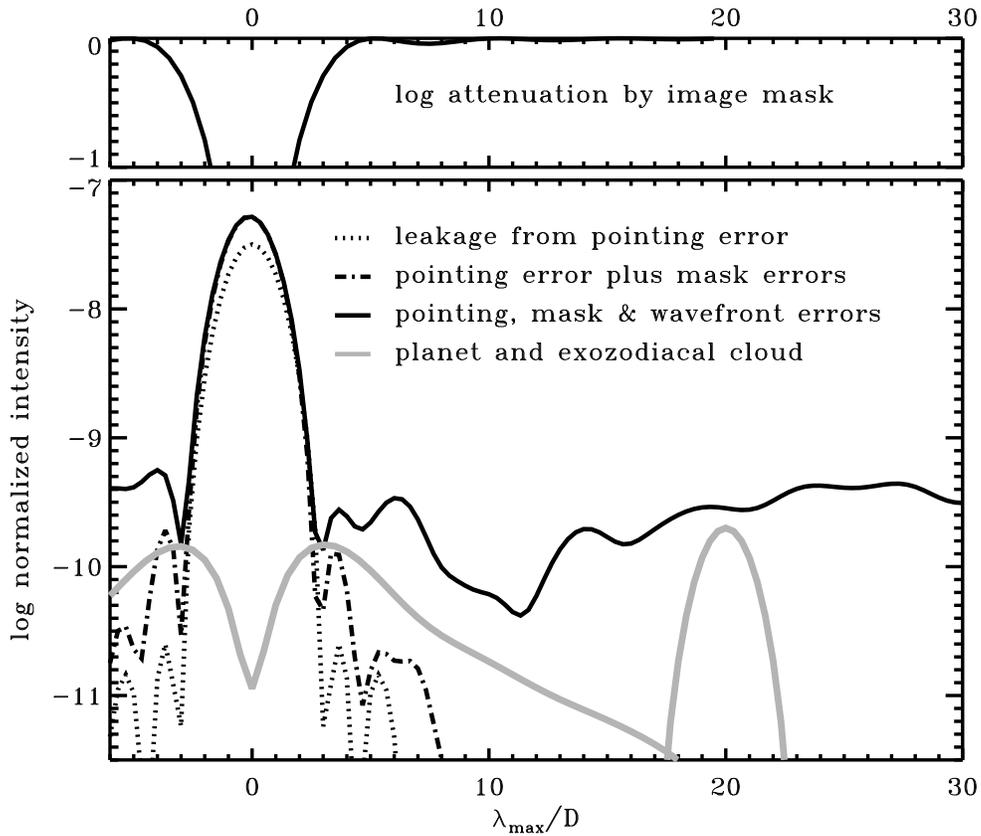}
\vspace{-0.8in}
\caption{Broadband simulation of images
produced by a coronagraphic TPF (Kuchner \& Spergel 2003).  The dotted
curve shows leakage due to pointing error.  The dot-dash curve adds
errors to the shape of the binary image mask.  The solid curve adds
amplitude and phase errors to the incoming wavefront.  The grey curves
show images of a planet (at $20 \lambda/D$) with relative flux $2
\times 10^{-10}$ and a $1 \times$ solar exozodiacal cloud 5 pc distant
assuming a 4m diameter primary.  The upper panel shows the attenuation
caused by the image mask.}
\end{figure}

Unlike the diffracted light problem, the scattered light problem will
not be considered completely solved until a TPF coronagraph has
successfully operated in space.  However, recent laboratory work
suggests that this goal lies within reach. Experiments using a
Xinetics deformable mirror at the High Contrast Imaging Testbed at JPL
have demonstrated active mirror control to the 0.25 \AA\ rms level
(0.5 \AA\ in wavefront phase) over mid-spatial frequencies (Trauger et
al. 2002a).  A pair of such mirrors can be combined to correct both
phase and amplitude errors over a reasonable bandpass.

Space can provide the necessary highly stable environment.
Thermal shields and other existing components appear to afford the
necessary mechanical and thermal isolation given an L2 orbit
and a mirror with a low thermal expansion coefficient,
though some subtle structural effects---like micro-snap---remain
unquantified.  Other areas that demand investigation are
the stability of optical coatings, and the sensitivity of the
designs to polarization effects.  These long-term stability concerns 
apply to mid-IR
interferometers too; an interferometer TPF requires hours to form an image
of a planet just as a coronagraphic TPF does, since the interferometer
must rotate to fill out its $(u,v)$ plane coverage.  

Coronagraphy merits resources like those nulling interferometry has
enjoyed for technology development and mission studies. The race
for choosing a design for Terrestrial Planet Finder has led to a
contest over which kind of high-dynamic range device is easier to
manufacture and implement.  Interferometry has a head start.  But for
a first mission to directly image planets, coronagraphy may have a
shorter way to go.

\section{Conclusion}

A 10~m class space coronagraph (Beichman et al. 2002) could complete
the originally mandated TPF survey of 150 F, G, and K stars.  A
smaller telescope (Brown et al. 2002)---perhaps one with an elliptical
6 by 2.7 m primary---probably suits the present TPF timeline.  A 2~m
class optical coronagraph in space (Trauger et al. 2002b) could
directly image Jupiter analogs and extrasolar planets too far from
their stars to detect by astrometric or precise-Doppler monitoring,
and test the new technologies needed for directly imaging extrasolar
terrestrial planets.

Both coronagraphy and interferometry require technological development
before either technique can directly detect extrasolar terrestrial
planets.  Perhaps an affordable mid-IR nulling interferometer could
also meet some of these scientific goals in the near future, and
ultimately, follow-up missions to TPF will demand the angular
resolution of a free-flying interferometer.  These follow-up missions,
however, might combine coronagraph and interferometer technologies.
And in the meantime, coronagraphy may offer a better way to find and
characterize and the first extrasolar analogs of Earth.

\acknowledgments

Thanks to Jeremy Kasdin, Steve Kilston, Charley Noecker and Wes Traub 
for close readings. This work was performed in part under contract with
the Jet Propulsion Laboratory (JPL) through the Michelson Fellowship program funded by 
NASA as an element of the Planet Finder Program.  JPL
is managed for NASA by the California Institute of Technology.

\end{document}